\documentclass{osa-article}

\journal{oe}

\articletype{Research Article}
\usepackage{cite} 
\usepackage{subfigure,enumerate}
\usepackage{amsmath}
\usepackage{booktabs} 
\usepackage{makecell}
\usepackage{colortbl}
\usepackage{setspace,arydshln}
\begin{document}

\title{Optimization of light fields in ghost imaging using dictionary learning}

\author{Chenyu Hu,\authormark{1,2} Zhishen Tong,\authormark{1,2} Zhentao Liu,\authormark{1} Zengfeng Huang,\authormark{3} Jian Wang\authormark{3,*} and Shensheng Han\authormark{1,2}}

\address{\authormark{1}Key Laboratory for Quantum Optics and Center for Cold Atom Physics of CAS, Shanghai Institute of Optics and Fine Mechanics, Chinese Academy of Sciences, Shanghai 201800, China\\
\authormark{2}Center of Materials Science and Optoelectronics Engineering, University of Chinese Academy of Sciences, Beijing 100049, China\\
\authormark{3}School of Data Science, Fudan University, Shanghai 200433, China}

\email{\authormark{*}jian\_wang@fudan.edu.cn} 



\begin{abstract}
Ghost imaging (GI) is a novel imaging technique based on the second-order correlation of light fields. Due to limited number of samplings in practice,
%
%
traditional GI methods often reconstruct objects with unsatisfactory quality. To improve the imaging results, many reconstruction methods have been developed, yet the reconstruction quality is still fundamentally restricted by the modulated light fields.
In this paper, we propose to improve the imaging quality of GI
by optimizing the light fields, which is realized via
matrix optimization for a learned dictionary incorporating the sparsity prior of objects. A closed-form solution of the sampling matrix, which enables successive sampling, is derived. Through simulation and experimental results, it is shown that
the proposed scheme leads to better imaging quality compared to the state-of-the-art optimization methods for light fields, especially at a low sampling rate.
%
%
\end{abstract}

\section{Introduction}
As a novel technique for optical imaging, ghost imaging (GI) was initially implemented with quantum-entangled photons two decades ago~\cite{pittman1995optical,strekalov1995observation}. In recent years, owing to its realization with thermal light and other new sources~\cite{bennink2002two,gatti2004ghost,cheng2004incoherent,zhang2005correlated,khakimov2016ghost,li2018electron},  GI has gained new attention and developed applications in various imaging areas, such as remote sensing~\cite{malik2012quantum,zhao2012ghost}, imaging through scattering media~\cite{Gong2011Correlated,Bina2013Backscattering}, spectral imaging~\cite{wang2016hyperspectral,liu2016spectral}, photon-limited imaging~\cite{morris2015imaging,liu2018fast} and X-ray imaging~\cite{pelliccia2016experimental,yu2016fourier}. 
\par Different from conventional imaging techniques that are based on the first-order correlation of light fields, GI extracts information of an object by calculating the second-order correlation between the light fields of the reference and the object arms~\cite{gatti2004ghost,cheng2004incoherent}. 
Theoretically, calculation of the second-order correlation requires infinite number of samplings of the light fields at both arms. In practice, however, the number of samplings is always finite, which often leads to reconstructed images of degraded signal-to-noise ratio (SNR)~\cite{xia2008experimental,erkmen2009signal,ferri2010differential}. To address this issue, much effort has been made in designing more effective reconstruction methods for GI.
On the one hand, approaches improving the second-order correlation have been proposed, which can increase the SNR of reconstructed images with theoretical guarantees~\cite{ferri2010differential,gong2010method,brida2010experimental,sun2012normalized}. On the other hand, by exploiting sparsity of the objects' images in transform bases~(e.g., wavelets~\cite{Antonini1992Image}), methods built upon the compressed sensing~(CS) theory~\cite{donoho2006compressed,candes2005decoding,candes2006near} have also been developed~\cite{katz2009compressive, han2018review}. In general, the CS based methods have superior performance over those relying on the second-order correlation, especially for imaging smooth objects~\cite{gong2013experimental}.
\par While improving the reconstruction methods has greatly promoted the practical applications of GI, there has been increasing evidence that the reconstruction quality of GI may be fundamentally restricted by the sampling efficiency~\cite{chen2014application,khamoushi2015sinusoidal}, i.e., how well information of objects is acquired in the samplings. To enable a satisfactory reconstruction from limited number of samplings in GI, a natural way is to enhance the sampling efficiency. In fact, this can be realized by optimizing the light fields of GI; see~\cite{Li2013Mutual,chen2014application,khamoushi2015sinusoidal,xu2015optimization} and the references therein.  
%
%
Considering an orthogonal sparsifying basis, Xu \emph{et al.}~\cite{xu2015optimization} optimized the sampling matrix in order for their product, so called the {\it equivalent sampling matrix},  to have the minimum mutual coherence, which results in much refinement of the imaging quality. 
%
%
Though orthogonal basis is widely suitable for sparse representation of natural images, for images from a specific category, dictionary learning~\cite{olshausen1996natural,aharon2006rm} usually produces much sparser representation coefficients, suggesting room for further improvements of the reconstruction quality. Motivated by this, in this paper we propose  to optimize the light fields of GI for a sparsifying basis obtained via dictionary learning. By minimizing the mutual coherence of the equivalent sampling matrix,  the proposed scheme enhances the sampling efficiency and thus achieves an improved reconstruction quality. In comparison  with the state-of-the-art optimization methods for  light fields in GI, the superiority of our scheme is confirmed via both simulations and experiments.
The main advantages of the proposed scheme is summarized as follows:

\begin{itemize}

\item Inspired from some previous researches in CS~\cite{elad2007optimized,abolghasemi2012gradient}, we formulate the problem of minimizing the mutual coherence of the equivalent sampling matrix in GI as a Frobenius-norm minimization problem, which yields a closed-form solution that depends on the sparsifying basis only. To the best of our knowledge, the suggested solution of the light fields is the first closed-form result in the GI optimization field.

\item The proposed scheme enables successive samplings. In GI, successive samplings means that when more samplings are available (or needed), one can simply augment new rows to the currently optimized sampling matrix in order to form a new one, without the need to perform additional optimization over the entire matrix. Such feature can bring great convenience to the practical applications of GI and was not addressed in previous works. 

%
%
%
%
%
%
%
%
%
%
%
%
%
\end{itemize} 

%
%
%
%
%
%
%
%
%
%
%
%
%

It is worth  mentioning that matrix optimization based on dictionary learning
has also been studied in the CS literature, see, e.g.,~\cite{elad2007optimized,abolghasemi2012gradient,duarte2009learning}. However, the optimizations in~\cite{elad2007optimized,abolghasemi2012gradient} were carried out over sampling matrices of fixed sizes, which  does not allow successive samplings. 
%
Although Duarte's method~\cite{duarte2009learning} dealt with the matrix optimization problem of alterable sampling size, it is also not compatible to GI  because of the demanding quantization accuracy. Moreover, those methods all fail to cope with the non-negative nature of sampling matrices in GI. 

\section{The Proposed Scheme}
The detection process in GI can be approximately formulated as~\cite{han2018review}
\begin{equation}
{\bf{y}} = {\bf{\Phi x}} + {\bf{n}},
\label{eq: CS GI}
\end{equation}
where ${\bf{y}} \in {\mathcal{R}}^M$ stands for the signal measured by the detector in the object arm,  ${\bf{\Phi}} \in {\mathcal{R}}^{M \times N}$ is the { sampling matrix} consisting of the light-field intensity distribution recorded by the detector in the reference arm, ${\bf{x}} \in {\mathcal{R}}^N$ signifies the object's information to be retrieved, and ${\bf{n}}$ denotes the detection noise. 
Let ${\bf \Psi}$ be the sparsifying basis obtained via dictionary learning, in which ${\bf{x}}$ can be sparsely represented as $\mathbf{x} = \mathbf{\Psi z}$, where $\mathbf{z}$ is the sparse coefficient vector. Also, consider the equivalent sensing matrix ${\bf{D}} := {\bf{\Phi \Psi}}$. Then, \eqref{eq: CS GI} can be rewritten as 
\begin{equation}
{\bf{y}} = {\bf{D z}} + {\bf{n}},
\label{eq: CS GI2}
\end{equation}
Evidences from the CS theory have revealed that a matrix $\mathbf{D}$ well preserving information of the sparse vector $\mathbf{z}$ guarantees a faithful reconstruction~\cite{donoho2006compressed,candes2005decoding,candes2006near}. 
As a powerful measure of information preservation, the mutual coherence $\mu(\mathbf{D})$ characterizes how incoherent each column pairs in $\mathbf{D}$ are~\cite{donoho2003optimally,donoho2006compressed}, namely, 
\begin{equation}
\mu \left( {\bf{D}} \right) = \mathop {\max} \limits_{1 \leq i < j \leq K} \frac{\big| \big\langle \mathbf{d}_i, \mathbf{d}_j \big \rangle  \big|} {\big\| \mathbf{d}_i \big\|_2  \big\| \mathbf{d}_j \big\|_2}
\end{equation}
with ${\bf{d}}_i$ being the $i$-th column of ${\bf{D}}$, $K$ the number of columns in $\bf{D}$ and $\|\cdot \|_2$ the $\ell_2$-norm.
For its simplicity and ease of computation, the mutual coherence $\mu\left( {\bf{D}} \right)$ has been widely used to describe the performance guarantees of CS reconstruction algorithms. For example, exact recovery of sparse signals via orthogonal matching pursuit (OMP)~\cite{pati1993orthogonal} is ensured by
$
\mu \left( {\bf{D}} \right) < \frac{1}{{2k - 1}}
$~\cite{tropp2004greed},
where $k$ is the sparsity level of input signals. In this work, with the aim of enhancing the sampling efficiency, we employ $\mu(\mathbf{D})$ as the objective function to be minimized in our optimization scheme. In particular, 
our proposed scheme consists of the following two main steps: 
\begin{itemize}
\item Firstly, an over-complete dictionary $\bf{\Psi}$ is learned from a collection of images, under the constraint that its first column has identical entries $N^{-1/2}$, while each of the other columns has entries summing to zero. Specifically, given ${\bf{X}} = [ {\bf x}^{(1)},{\bf x}^{(2)}, \cdots ,{\bf x}^{(K)} ] \in {\mathcal{R}^{N \times L}}$, in which each column is a reshaped vector of the training image sample, the sparsifying dictionary ${\bf \Psi} \in \mathcal{R}^{N \times K}$ is learned by solving the following problem:
\begin{eqnarray}
&\min_{\bf \Psi, Z} & \left\| {{\bf{X}} - {\bf{\Psi Z}}} \right\|_F^2 \nonumber \\
&{\rm{subject \ to}} &{\bf{\Psi}}_{11} = \cdots = {\bf{\Psi}}_{N1} = N^{-1/2}, \nonumber \\
&&  \|{\bf{z}}_i\|_0 \le T_0,~i = 1, \cdots, L,
\label{eq: DicLearning}
\end{eqnarray}
where $\|\cdot\|_F$ and $\|\cdot\|_0$ are the Frobenius- and $\ell_0$-norm, respectively, ${\bf{Z}} = \big[ {\bf z}_1,{\bf z}_2, \cdots ,{\bf z}_L \big] \in \mathcal{R}^{K \times L}$ represents the sparse coefficient matrix of training images, and $T_0$ denotes the predetermined sparsity level of vectors ${\bf{z}}_i$. In this work, we shall employ $K$-SVD as a representative  method to perform the dictionary learning task, which results in simultaneous sparse representation of input images in the learned dictionary~$\bf{\Psi}$. Readers are referred to~\cite{aharon2006rm} for more details of the $K$-SVD method. 

\item Secondly, the sampling matrix $\bf{\Phi}$ is optimized by minimizing the mutual coherence of the equivalent sampling matrix $\bf{D}$. Put formally, 
\begin{equation}
\mathop {\min }\limits_{\bf{\Phi }} \mu \left( \mathbf{D} \right) ~~\ \ {\rm{subject \ to}} ~\ \ \ {{\bf{\Phi }}_{ij}} \ge 0~\text{and}~{\bf{D}} = \bf{\Phi\Psi}.	\label{func:coherence optimization}
\end{equation}
The non-negative constraint $\mathbf{\Phi}_{ij} \geq 0$ is imposed due to the fact that the intensity of light fields is always non-negative. 
\end{itemize}
\par We now proceed to solve the optimization problem in~(\ref{func:coherence optimization}). Without loss of generality, assume that matrix $\mathbf{D}$ has $\ell_2$-normalized columns, that is, $\|\mathbf{d}_i\|_2 = 1$ for $i = 1, \cdots, K$. Then,  
\begin{equation}
\mu(\mathbf{D}) =  \max_{1 \leq i < j \leq K} {| \langle \mathbf{d}_i, \mathbf{d}_j \rangle |}.
\end{equation}
To optimize $\mu(\mathbf{D})$, it suffices to minimize the off-diagonal entries of the Gram matrix $\mathbf{D}^{\top} \mathbf{D}$, each of which corresponds to the coherence between two different columns in $\mathbf{D}$ (i.e., $
(\mathbf{D}^{\top} \mathbf{D})_{ij} =  | \big\langle \mathbf{d}_i, \mathbf{d}_j \big \rangle |$, $i \neq j$). In particular, we would like 
%
 %
the Gram matrix to be as close to the identity matrix as possible, namely, $\mathbf{\Psi}^{\top} \mathbf{\Phi}^{\top} \mathbf{\Phi \Psi} \approx \bf{I}.$ Since replacing the identity matrix with ${\bf{\Psi}}^{\top}{\bf{\Psi}}$ yields a sampling matrix robust to the sparse representation error of images~\cite{cleju2014optimized}, we propose to optimize $\mathbf{\Phi}$ via
\begin{equation}
\mathop {\min }\limits_{\bf{\Phi }} \big \| {{{\bf{\Psi }}^{\top}}{{\bf{\Phi }}^{\top}}{\bf{\Phi \Psi }} - {{\bf{\Psi }}^{\top}}{\bf{\Psi }}} \big\|_F^2.
\label{eq: OriginalProblem}
\end{equation} 

By multiplying $\bf{\Psi}$ and ${\bf{\Psi}}^{\top}$ on the left- and right-hand sides of both terms inside the Frobenius norm, respectively, one has
\begin{equation}\label{eq: OriginalProblem1}
\mathop {\min }\limits_{\bf{\Phi }} \big \| {{\bf{\Psi }}  {{\bf{\Psi }}^{\top}} {{\bf{\Phi }}^{\top}}{\bf{\Phi \Psi }}  {{\bf{\Psi }}^{\top}} - {\bf{\Psi }}  {{\bf{\Psi }}^{\top}}{\bf{\Psi }}  {{\bf{\Psi }}^{\top}}} \big\|_F^2.
\end{equation}
After substituting ${\bf{\Psi }}{{\bf{\Psi }}^{\top}}$ with its eigenvalue decomposition ${\bf{V}}{\bf{\Lambda }}{{\bf{V}}^{\top}}$, and also denoting ${\bf{W}} := {\bf{\Lambda }}{{\bf{V}}^{\top}}{{\bf{\Phi }}^{\top}}$,  (\ref{eq: OriginalProblem1}) can be rewritten as 
\begin{equation}\min_{\bf{W}} \big\| {{\bf{VW}}{{\bf{W}}^{\top}}{{\bf{V}}^{\top}} - {\bf{V}}{\bf \Lambda}^2 {{\bf{V}}^{\top}}} \big\|_F^2,
\end{equation}
or equivalently,
\begin{equation}
\mathop {{\rm{min}}}\limits_{\bf{W}} \Bigg\| {{{\bf{\Lambda }}^2} - \sum\limits_{i = 1}^M {{{\bf{w}}_i}{\bf{w}}_i^{\top}} } \Bigg\|_F^2~\text{where}~{\bf{W}} = \left[ {{{\bf{w}}_1}, \cdots ,{{\bf{w}}_M}} \right].
\label{eq: Transformed Problem}
\end{equation}
Denoting ${\bf{\Lambda }} = \left[ {{{\bf{r}}_1}, \cdots ,{{\bf{r}}_N}} \right]$, (\ref{eq: Transformed Problem}) further becomes
\begin{equation}
\mathop {{\rm{min}}}\limits_{\bf{W}} \Bigg\| {\sum\limits_{j = 1}^N {{{\bf{r}}_j}{\bf{r}}_j^{\top}}  - \sum\limits_{i = 1}^M {{{\bf{w}}_i}{\bf{w}}_i^{\top}} } \Bigg\|_F^2.
\label{eq: FinalProblem}
\end{equation}

Clearly problem~(\ref{eq: FinalProblem}) has the solution 
$
\widehat{{\bf{W}}} = {{\bf{\Lambda }}_1^{\top}},
$
where ${\bf{\Lambda}}_1$ is the matrix consisting of the first $M$ columns of $\bf{\Lambda}$, which is obtained by setting ${{\bf{w}}_k} = {{\bf{r}}_k}$, $k = 1, \cdots, M$. Recalling that ${\bf{W}} := {\bf{\Lambda }}{{\bf{V}}^{\top}}{{\bf{\Phi }}^{\top}}$, the optimized sampling matrix $\bf{\Phi}$ can be simply calculated as
\begin{equation}
{\widehat {\bf{\Phi }} = {\widehat {\bf{W}}^{\rm{T}}}{\left( {{{\bf{\Lambda }}^{ - 1}}} \right)^{\rm{T}}}{{\bf{V}}^{\rm{T}}} = {{\bf{\Lambda }}_1}{\left( {{{\bf{\Lambda }}^{ - 1}}} \right)^{\rm{T}}}{{\bf{V}}^{\rm{T}}} = \left[ {\begin{array}{*{20}{c}}
{{{\bf{I}}_{M \times M}}}&{\bf{0}}
\end{array}} \right]\left[ {\begin{array}{*{20}{c}}
{{\bf{V}}_1^{\rm{T}}}\\
{{\bf{V}}_2^{\rm{T}}}
\end{array}} \right] = {\bf{V}}_1^{\rm{T}}},
\label{eq: OptimizedMatrix}
\end{equation}
\noindent where matrix ${\bf{V}}_1$ consists of the first $M$ columns of $\bf{V}$. When more samplings become available, interestingly, it suffices to update $\widehat{\bf{\Phi}}$ by augmenting more rows of $\mathbf{V}^{\top}$ to the previous one, thereby enabling a successive sampling. As aforementioned, the feature of successive sampling is of vital importance to the practical applications of GI.

\par Furthermore, due to the fact that the intensity of light fields is always non-negative, additional treatments are needed to make sure that elements of the sampling matrix are non-negative (NN). To the end, we propose a NN lifting, which adds a constant matrix to the optimized matrix $\widehat{\bf{\Phi}}$ in~(\ref{eq: OptimizedMatrix}) as
\begin{equation}
\widehat{\mathbf{D}} = \big( {\widehat{\bf{\Phi }} + c{\bf{1}}_{M \times N}} \big){\bf{\Psi }},
\end{equation}
where $\mathbf{1}_{M \times N}$ is an $M$-by-$N$ matrix with entries being ones and 
\begin{equation}
c : =
\begin{cases}
- \min_{i,j } \widehat{\mathbf{\Phi}}_{ij}                       & \text{if}~\min_{i,j } \widehat{\mathbf{\Phi}}_{ij} < 0, \\ 
0                      & \text{if}~\min_{i,j } \widehat{\mathbf{\Phi}}_{ij} \geq 0.
\end{cases} \label{eq:jjjjffff}
\end{equation}
As aforementioned, the first column of $\mathbf{\Psi}$ has identical entries and other columns have entries summing to zero. Thus, 
\begin{eqnarray} 
 \widehat{\bf{D}}
 &\hspace{-2mm} 
=&\hspace{-2mm} \widehat{\bf{\Phi }} {\bf{\Psi}}
+ c N^{-1/2} \big[ {{\bf{1}}_{M \times 1},\underbrace{{\bf{0}}, \ldots ,{\bf{0}}}_{M \times (K - 1)}} \big].
 \label{eq: SimMu} 
\end{eqnarray}
It can be noticed that after the NN lifting, $\widehat{\bf{D}}$ and $\widehat{\bf{\Phi }} {\bf{\Psi}}$ differ only in the first column. Nevertheless, the mutual coherence $\mu(\widehat{\mathbf{D}})$ is not much affected by the NN lifting, as confirmed by our extensive empirical test.

\section{Results}
\subsection{Simulations}
To evaluate the effectiveness of the proposed optimization scheme, both simulations and experiments are performed. MNIST handwritten digits of size $28 \times 28$ pixels~\cite{deng2012mnist} are chosen to be the imaging objects, and the dictionary $\bf{\Psi}$ is learned based on 20,000 digits randomly selected from the training set. 
\begin{figure}[t] 
\begin{center} 
\subfigure[]{\includegraphics[width=0.45\textwidth]{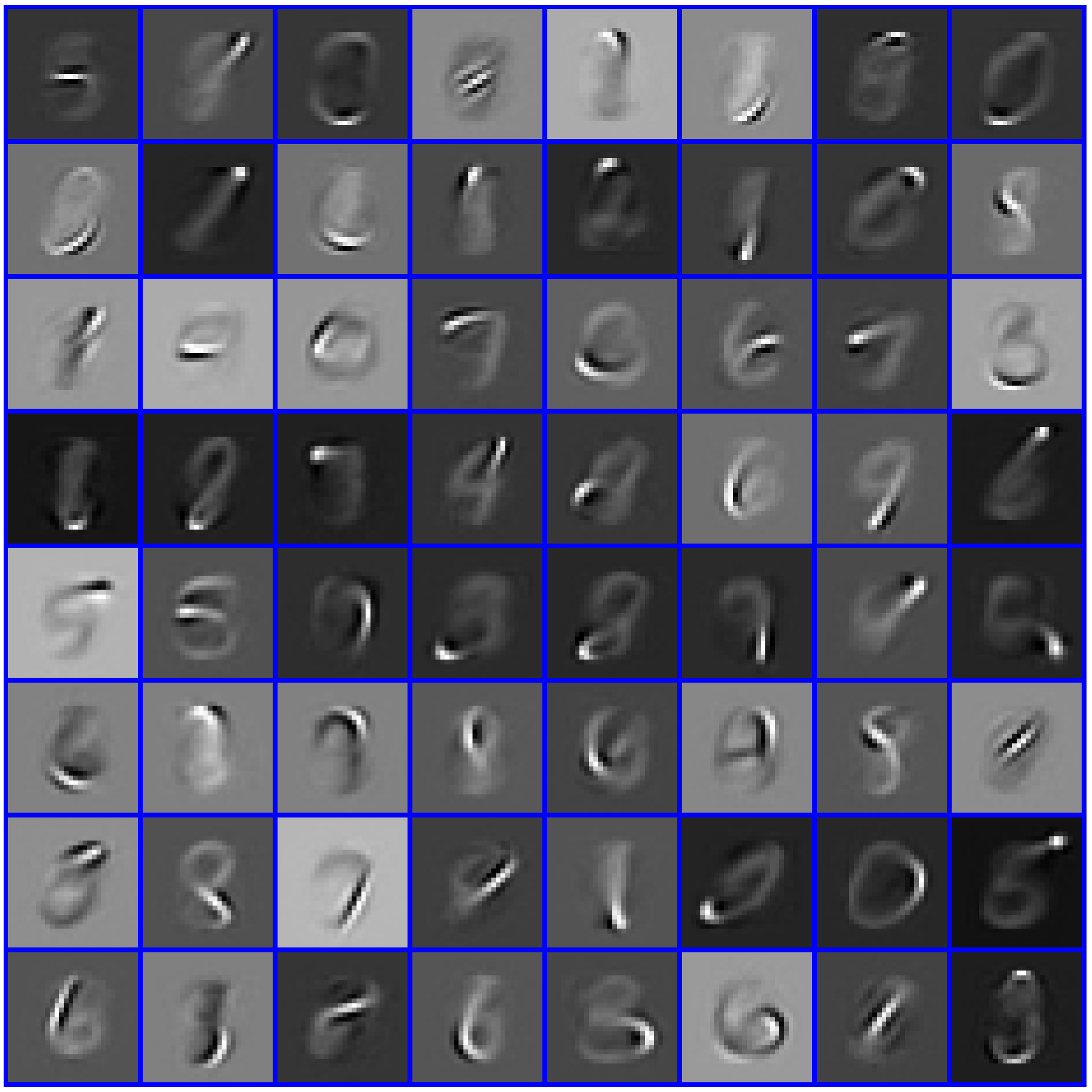}\label{fig: Atoms in the Dictionary}} 
\hspace{2mm}
\subfigure[]{\includegraphics[width=0.45\textwidth]{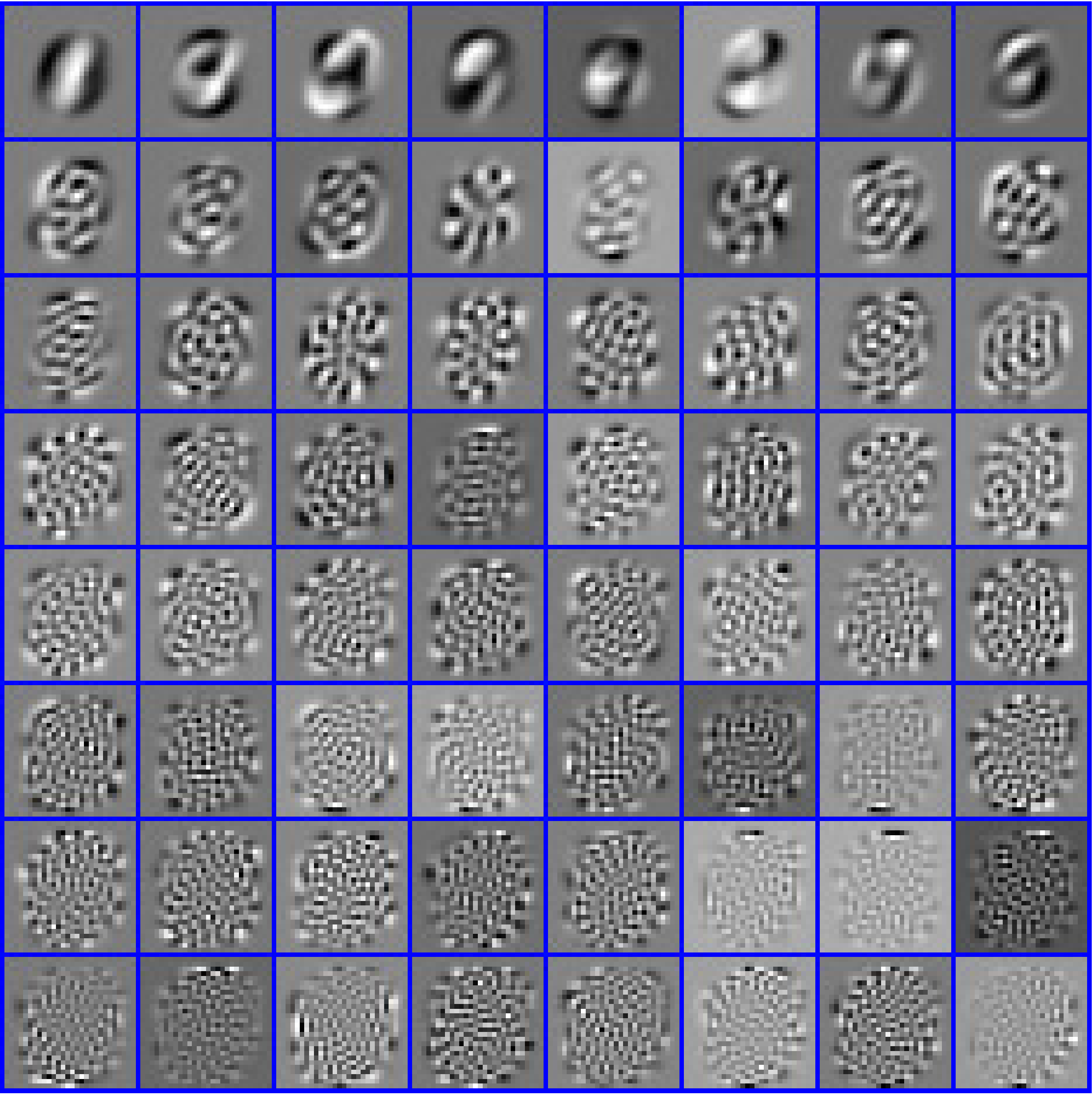}\label{fig: SamplingPatterns}} 
\end{center}
\caption{(Left) A subset of atoms in the learned dictionary $\mathbf{\Psi}$; (right) a subset of the optimized lgiht-field intensity distributions}  
\label{fig: Dic and SamMat}
\end{figure}

\begin{table}[h]
    \caption{A summary of test methods}
    \setlength{\arrayrulewidth}{.75pt}
    \centering
    \begin{tabular}{c|c|c|c}
    \hline
    \textbf{Method}                  & \textbf{Sampling matrix}          & \textbf{Dictionary}                  & $\mathbf{\Phi}_{ij} \geq 0$\\
    \hline  \hline    
    Proposed                         & Eq. (\ref{eq: OptimizedMatrix})   & $K$-SVD                 & NN-lifting\\
    \hline 
    Gaussian                         & Random Gaussian            & $K$-SVD                & NN-lifting\\
    \hline 
    Duarte~\cite{duarte2009learning} & Matrix optimization               & $K$-SVD                 & NN-lifting\\
    \hline 
    Xu~\cite{xu2015optimization}     & Matrix optimization               & DCT                                                & Zero-forcing\\
    \hline 
    NGI~\cite{sun2012normalized}     & Random Gaussian                   & None &   NN-lifting  \\
    \hline
    \end{tabular}
    \label{tab:my_label}
\end{table}

Moreover, the optimized sampling matrix $\widehat{\mathbf{\Phi}}$ is obtained from~(\ref{eq: OptimizedMatrix}), followed by the NN lifting. A subset of atoms in the learned dictionary $\mathbf{\Psi}$ and the optimized light-field intensity distributions $\widehat{\mathbf{\Phi}}$ are shown as Fig.~\ref{fig: Atoms in the Dictionary} and Fig.~\ref{fig: SamplingPatterns}, respectively.
For comparative purposes, our simulation includes other four methods: 1)~Gaussian method,  2)~Duarte's method~\cite{duarte2009learning}, 3)~Xu's method~\cite{xu2015optimization} and 4)~normalized GI (NGI) method~\cite{sun2012normalized}. {\textbf{Table~\ref{tab:my_label}}} gives a brief summary of the methods under test.
In the Gaussian method, the sampling matrices are random Gaussian matrices, whose entries are drawn independently from the standard Gaussian distribution ($\mathbf{\Phi}_{ij} \sim \mathcal{N}(0,1)$). To meet the NN constraint $\mathbf{\Phi}_{ij} \geq 0$ of GI,  the matrices $\mathbf{\Phi}$'s generated from the Gaussian and Duarte's methods are also inflicted with the NN lifting.
For the Gaussian, Duarte's and our proposed methods, the images are retrieved in two steps. Firstly, the sparse coefficient vector $\hat{\mathbf{z}}$ of image under the learned dictionary $\bf{\Psi}$ is obtained by solving the $\ell_0$-minimization problem:
\begin{equation}
\hat{\mathbf{z}} = \underset{\mathbf{z}}{\arg \min} \hspace{0.5mm} \|\widehat{\bf{D}} \mathbf{z} - \mathbf{y} \|_2^2~~~\text{subject~to}~~\|\mathbf{z}\|_0 \leq T_0
\label{eq: OMPRec}
\end{equation}
via the OMP algorithm~\cite{pati1993orthogonal}.
Secondly, the object's image is reconstructed as
\begin{equation}
\widehat{\bf{x}}={\bf{\Psi}}{\widehat{{\bf{z}}}}. \label{eq:17}
\end{equation}
For Xu's method, the Discrete Cosine Transform (DCT) basis is chosen as the orthogonal basis. And the images in methods 3) and 4) are reconstructed via approaches proposed in their corresponding references.

\begin{figure}[t] 
\begin{center} 
\subfigure[]{\includegraphics[width=0.85\textwidth]{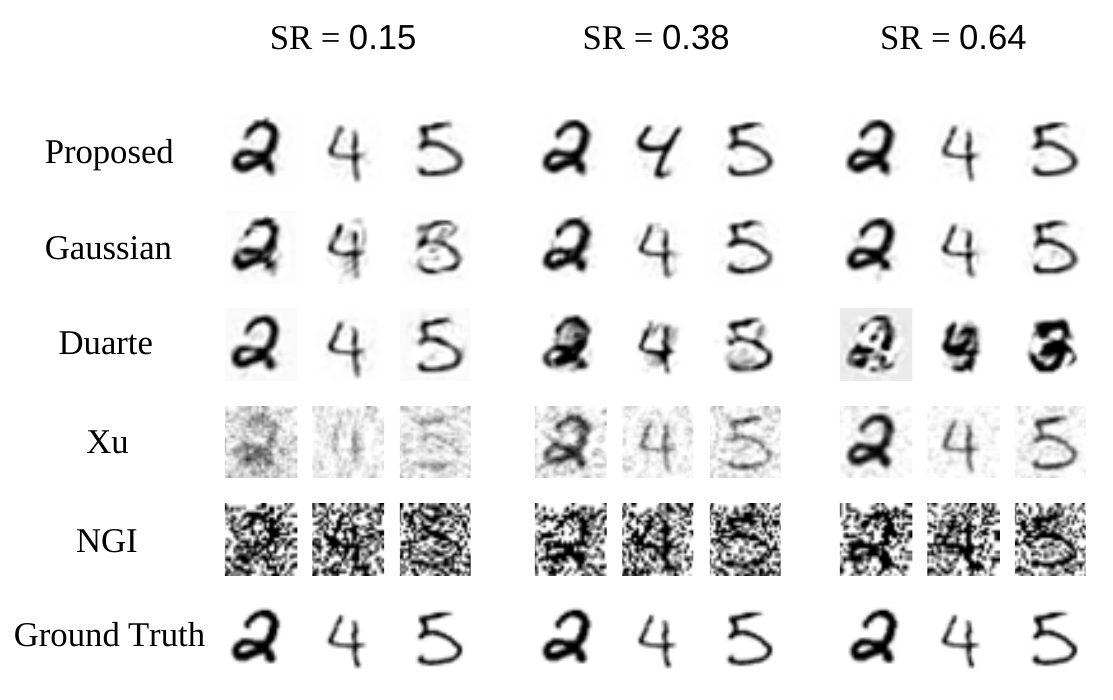}\label{fig: RecImg of Simulation1}} 
\subfigure[]{\includegraphics[width=0.495\textwidth]{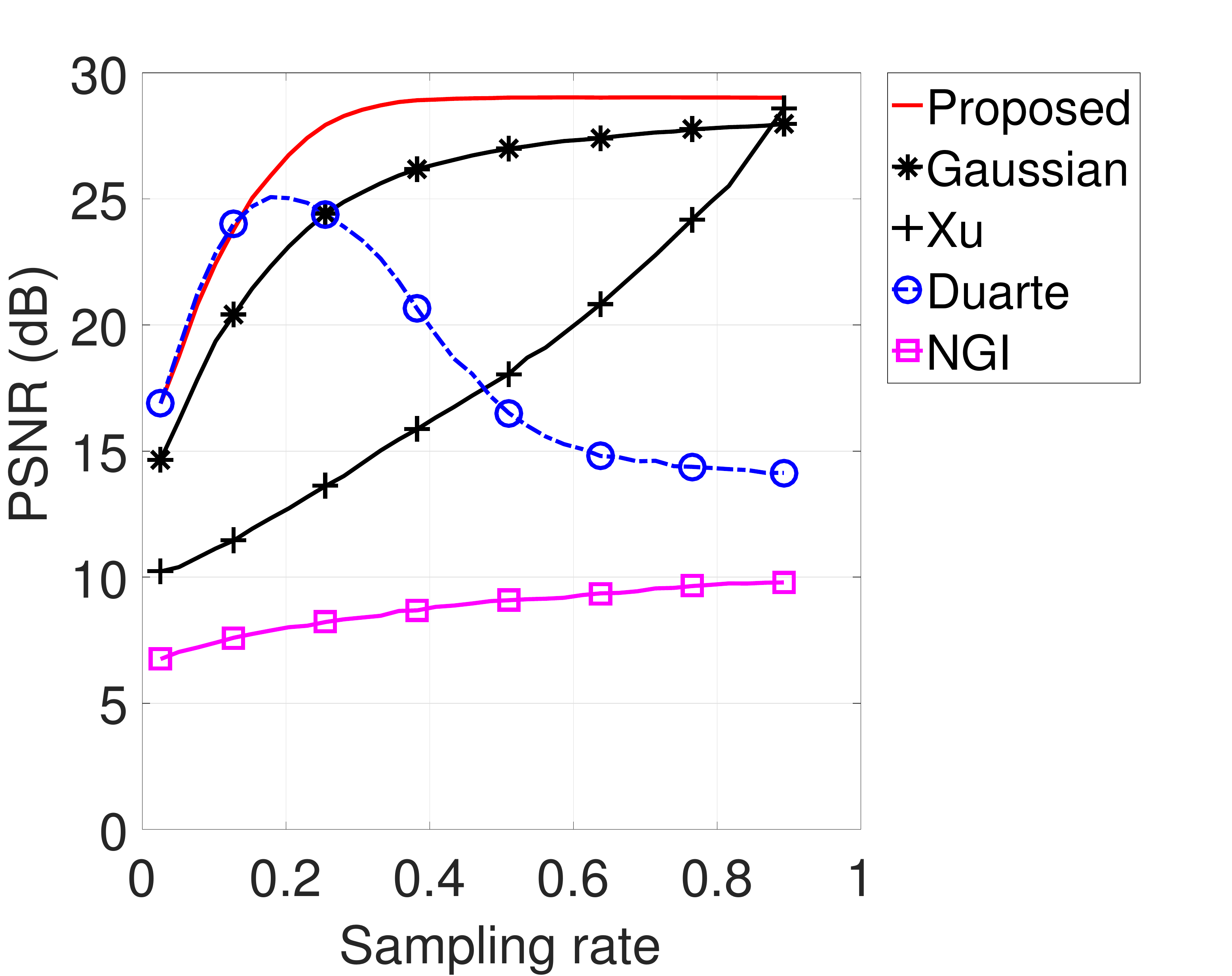}\label{fig: RecPSNR of Simulation1}}
\subfigure[]{\includegraphics[width=0.495\textwidth]{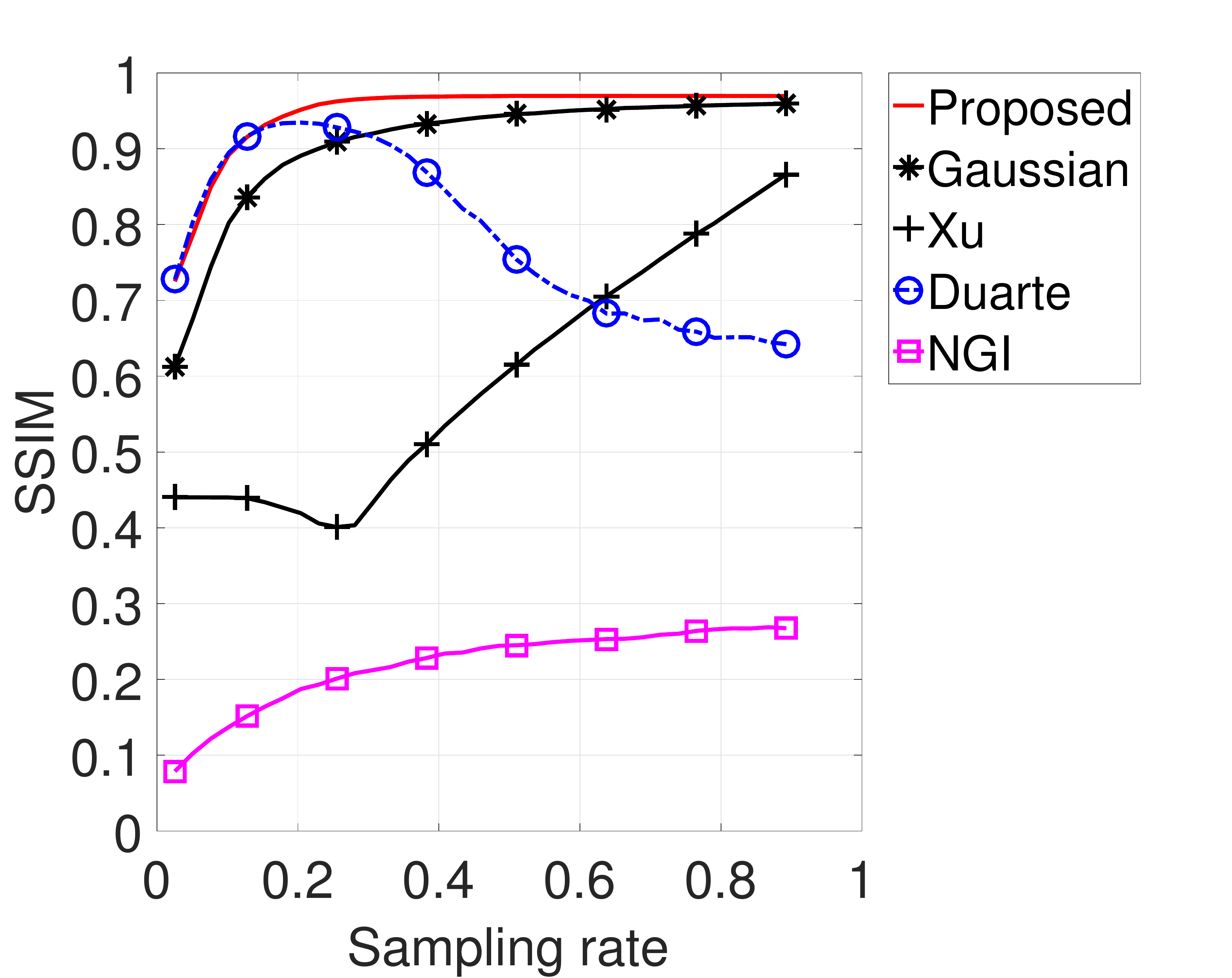}\label{fig: RecSSIM of Simulation1}}
\end{center}
\caption{Simulation results with sampling matrix of high accuracy. Fig. \ref{fig: RecImg of Simulation1} shows the reconstructed images reconstructed via different methods under different SR. Fig. \ref{fig: RecPSNR of Simulation1} and \ref{fig: RecSSIM of Simulation1} illustrate the PSNR and SSIM of reconstructed images via different methods as a function of SR, respectively.} 
\label{fig: Simulation1}  
\end{figure}

In our simulation, we first adopt matrices with entries of double-type in MATLAB. Fig.~\ref{fig: Simulation1} shows the simulation results of different methods.  
In Fig.~\ref{fig: RecImg of Simulation1}, the reconstructed images at the different sampling ratios~(SR's) (i.e., SR $=  0.10$, $0.20$, and $0.51$) are displayed, where the SR is computed by dividing the number of samplings by the number of image pixels.
Fig.~\ref{fig: RecPSNR of Simulation1} and~\ref{fig: RecSSIM of Simulation1} depict the peak signal-to-noise ratio (PSNR) and structural similarity (SSIM) index of the reconstructed images as functions of the SR, respectively.  Given the reference image $X$ and the reconstructed image $Y$, the PSNR and SSIM are defined as follows, 
\begin{subequations}
\begin{eqnarray}
&&{\rm{MSE}}\left( {X,Y} \right) = \frac{1}{{mn}}\sum\limits_{i=1}^m\sum\limits_{j=1}^n {{{\left[ {X\left( {i,j} \right) - Y\left( {i,j} \right)} \right]}^2}}, \\
&&{\rm{PSNR}}\left( {X,Y} \right) = 10 \hspace{0.4mm} {\log _{10}}\left[ {\frac{{{{B}^2}}}{{{\rm{MSE}}\left( {X,Y} \right)}}} \right], \\
&&{\rm{SSIM}}\left( {X,Y} \right) = \frac{{\left( {2{\mu _X}{\mu _Y} + {c_1}} \right)\left( {2{\sigma _{XY}} + {c_2}} \right)}}{{\left( {\mu _X^2 + \mu _Y^2 + {c_1}} \right)\left( {\sigma _X^2 + \sigma _Y^2 + {c_2}} \right)}},
\end{eqnarray}
\end{subequations}
where the pixel size of image is $m \times n$, $B$ denotes the dynamic range of image pixels, which takes the value $255$ in this paper, $(\mu _X, \sigma _X)$ and $(\mu _Y, \sigma _Y)$ are the means and variances of $X$ and $Y$, respectively,  $\sigma _{XY}$ is the covariance of $X$ and $Y$, $c_1 = (0.01B)^2$ and $c_2 = (0.03B)^2$.
These two metrics measure the difference and similarity between the reconstructed images and the original ones, respectively. For each SR under test, the PSNR and SSIM are averaged over 500 reconstructed digit images to plot the curves.  By comparing them, the reconstruction quality of different methods are compared empirically.

\begin{figure}[t]
\begin{center}
\subfigure[]{\includegraphics[width=0.85\textwidth]{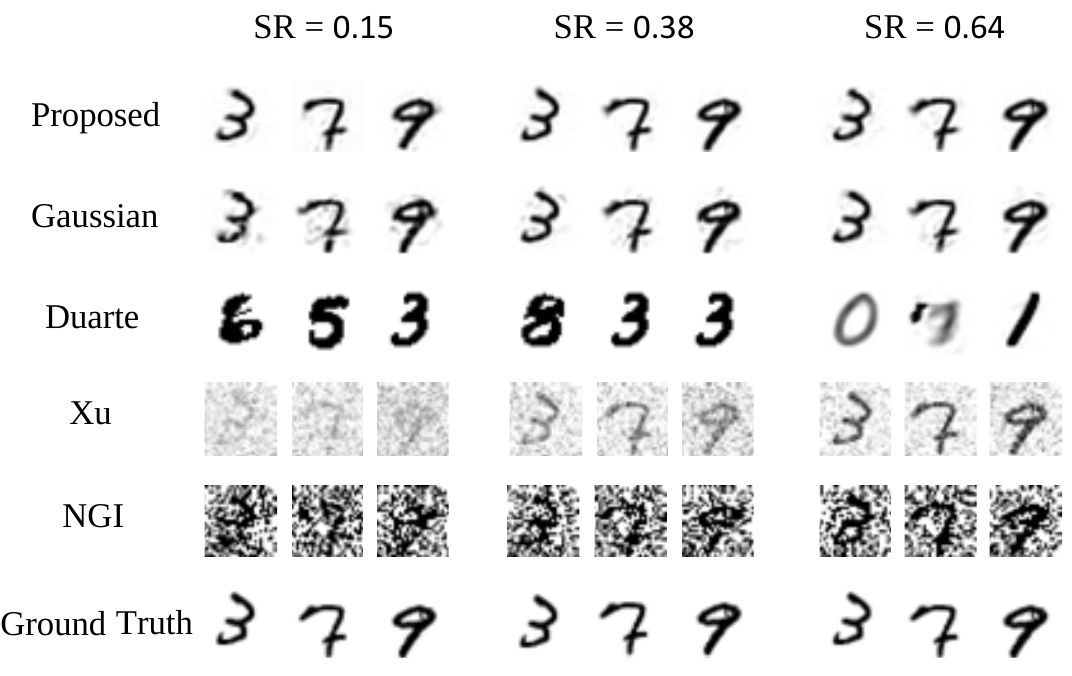}\label{fig: RecImg of Simulation2}} 
\subfigure[]{\includegraphics[width=0.495\textwidth]{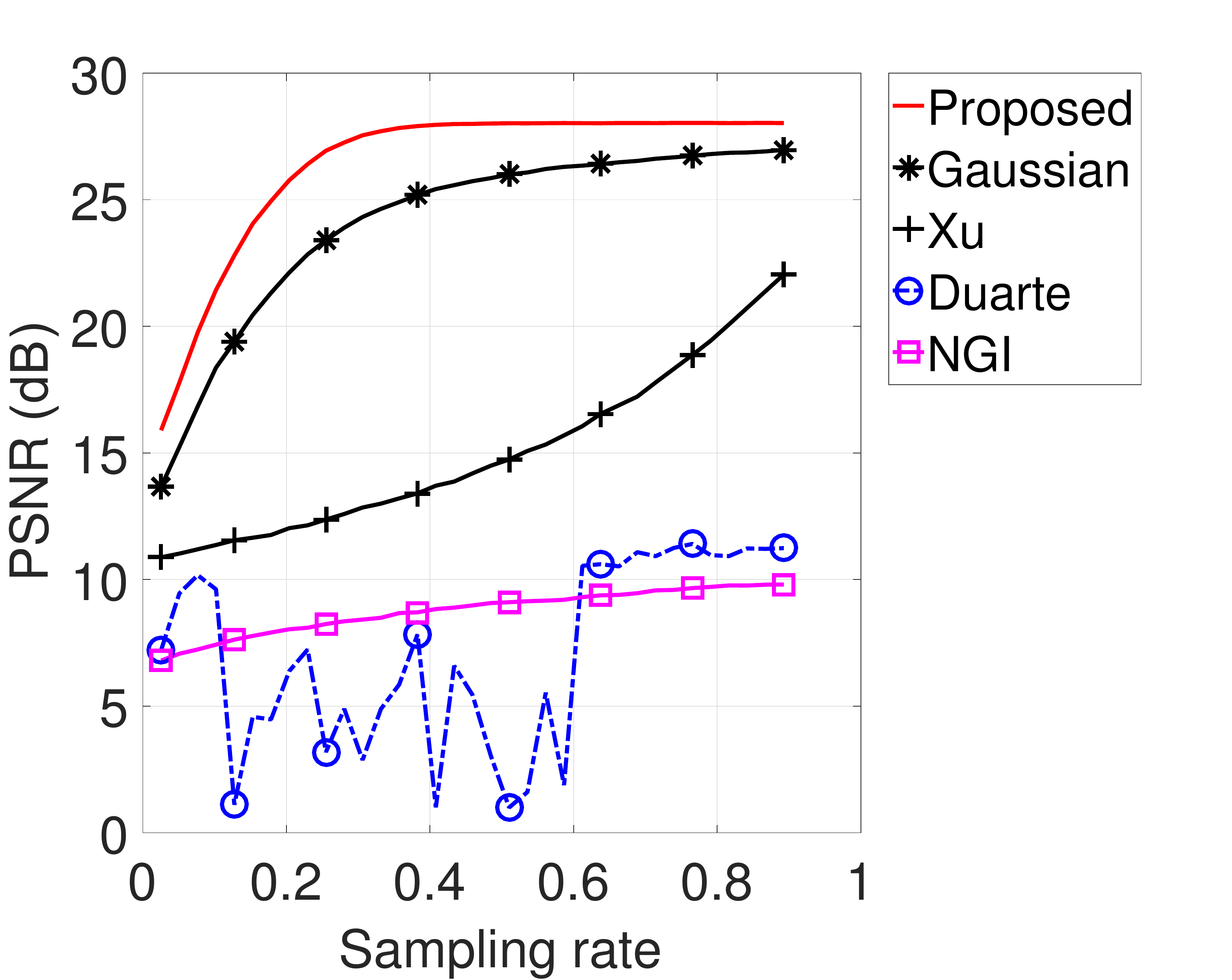}\label{fig: RecPSNR of Simulation2}}
\subfigure[]{\includegraphics[width=0.495\textwidth]{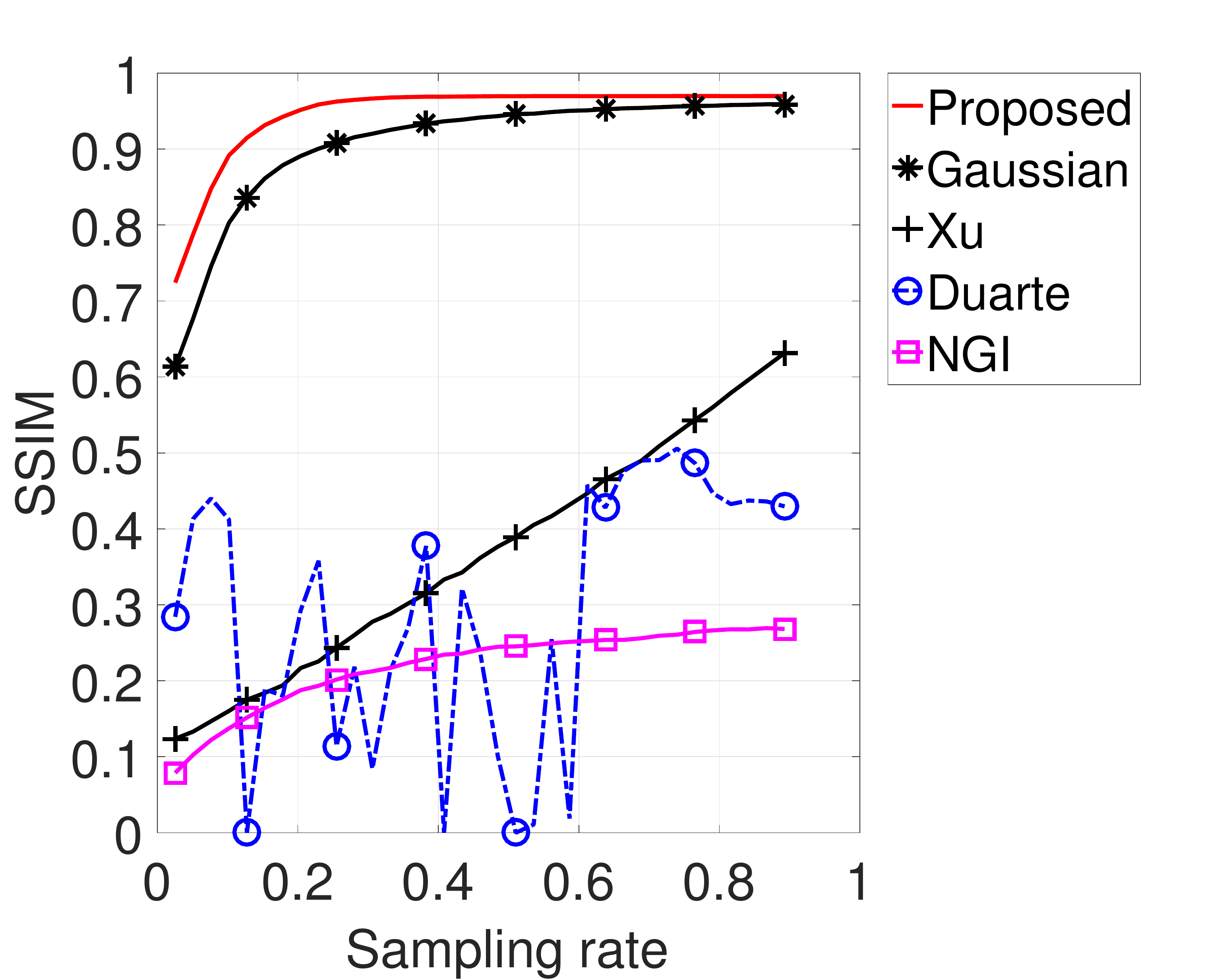}\label{fig: RecSSIM of Simulation2}}
\end{center}
\caption{Simulation results with $8$-bit quantized sampling matrix. Fig.~\ref{fig: RecImg of Simulation2} shows the reconstructed images via different methods under different SR. Fig.~\ref{fig: RecPSNR of Simulation2} and \ref{fig: RecSSIM of Simulation2} show PSNR and SSIM of the reconstructed images via different methods as a function of SR, respectively.}
\label{fig: Simulation2} 
\end{figure}

%



%

From Fig.~\ref{fig: Simulation1}, it can be observed that the reconstruction quality of the proposed scheme is uniformly better than the other methods under test. In particular, it achieves $2$dB to $4$dB gain of PSNR and up to $10$\% higher SSIM over the Gaussian method, owing to the optimized light fields. 
Compared to Xu's method~\cite{xu2015optimization}, the Gaussian method have a notable advantage in the low SR region, which gradually converges as the SR approaches one. The performance gap is mainly attributed to the utilization of dictionary learning that can better incorporate the sparsity prior of images. Among all test methods, the PSNR and SSIM of the NGI method lie in the lowest level. This is mainly due to the image noise, which often happens to the correlation-based reconstruction methods of GI, especially when the SR is low.
%
%
%
%
It can also be observed from Fig.~\ref{fig: Simulation1} that Duarte's method~\cite{duarte2009learning} performs comparably with the proposed method in the low SR region, but deteriorating dramatically when the SR increases. Such phenomenon seems unreasonable at first glance, but can be interpreted from the condition number perspective. 
To be specific, the sampling matrix ${\bf \Phi}$ of Duarte's method has larger condition number as the SR increases (see detailed explanations in Footnote 7 of~\cite{duarte2009learning}). Thus, when ${\bf \Phi}$ multiplies with the representation error ${\bf{e}}  = {\bf{x - \Psi z}}$, it could significantly amplify this error, and eventually degrade the reconstruction quality. Indeed, in this case one would need to reconstruct the sparse vector $\bf z$ from the samplings ${\bf{y}} = {\bf{\Phi x}} + {\bf{n}} = {\bf{\Phi }} {\bf{\Psi z}} + {\bf \Phi e} + {\bf{n}} = {\bf{D z}}  + ({\bf \Phi e} + {\bf{n}})$, which can be difficult since the largely amplified error ${\bf \Phi e}$ essentially becomes part of noise for the reconstruction.


\begin{figure}[t]
\centering
\includegraphics[width=0.75\textwidth]{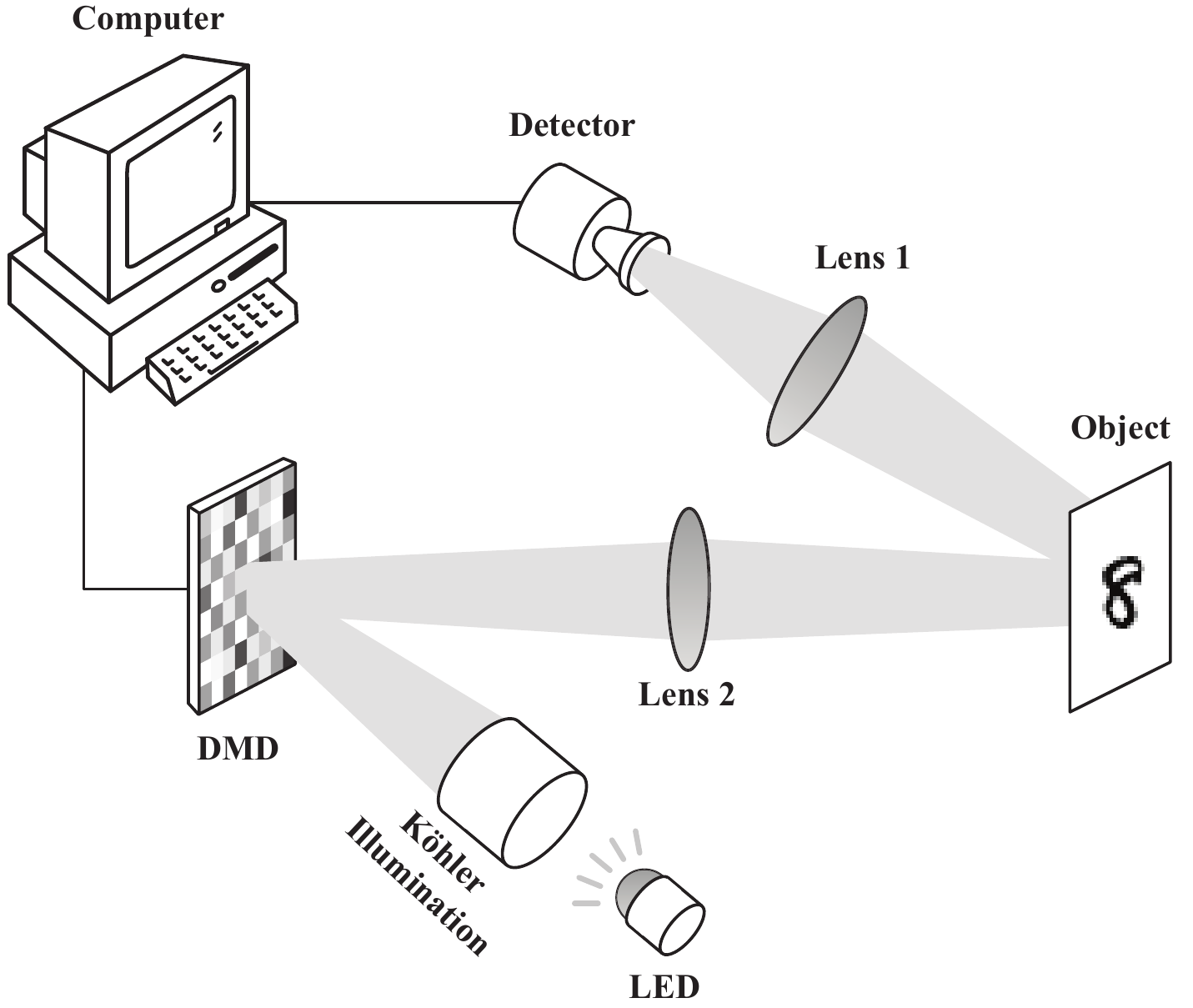}
\caption{Schematic diagram of experimental setup. Light-field patterns are generated by DMD and projected onto the object, afterwards collected by the detector.}
\label{fig: ExperimentalSetup} 
\end{figure}

\begin{figure}[t]
\begin{center}
\subfigure[]{\includegraphics[width=0.85\textwidth]{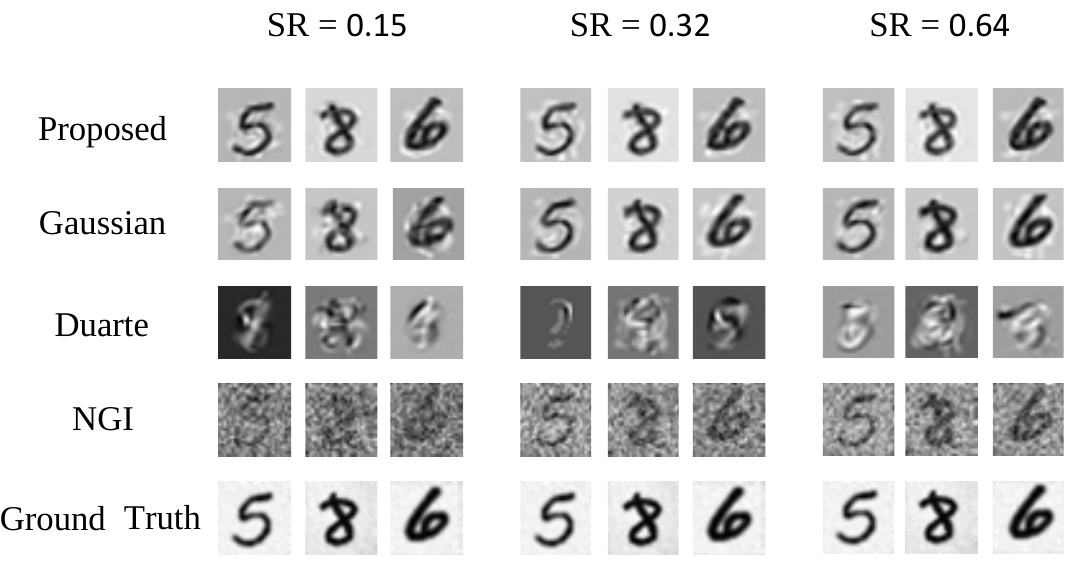} \label{fig: RecImg of Experiments}}
\subfigure[]{\includegraphics[width=0.45\textwidth]{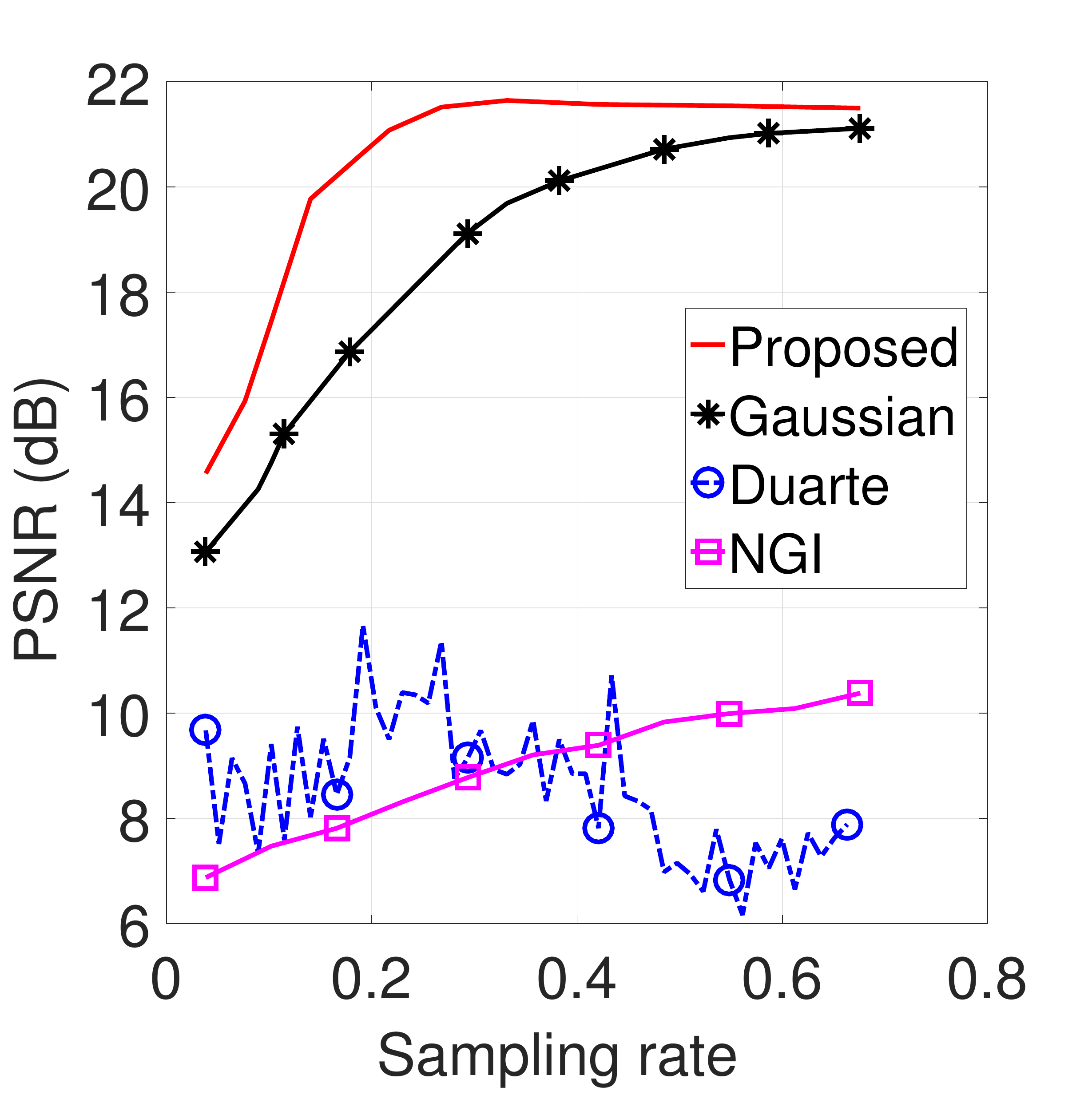} \label{fig: RecPSNR of Experiments}}
\hspace{8mm}
\subfigure[]{\includegraphics[width=0.45\textwidth]{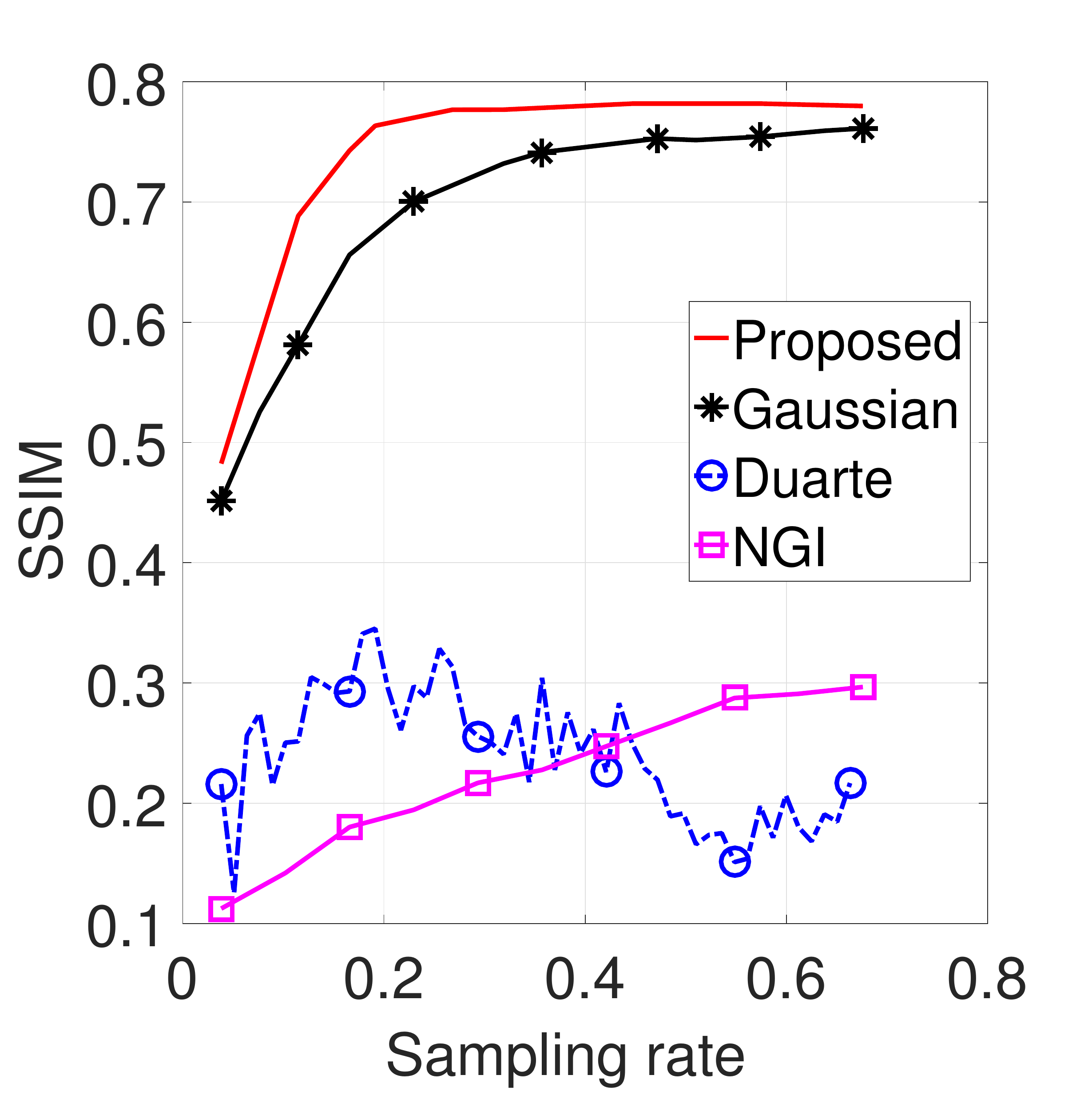} \label{fig: RecSSIM of Experiments}}
\end{center}
\caption{Experimental results. Fig.~\ref{fig: RecImg of Experiments} shows the reconstructed images via different methods under different SR. Fig.~\ref{fig: RecPSNR of Experiments} and~\ref{fig: RecSSIM of Experiments} show PSNR and SSIM of the reconstructed images via different methods as a function of SR, respectively.}
\label{fig: Experiments} 
\end{figure}


\par In practice, detectors measure the intensity signals with quantization, which means that $\bf{\Phi}$ is actually a quantized sampling matrix. Thus, we also simulate the case where the sampling matrices are quantized to $8$-bit of precision and plot the results in Fig.~\ref{fig: Simulation2}. Similarly, Fig.~\ref{fig: RecPSNR of Simulation2} and~\ref{fig: RecSSIM of Simulation2} show the curves of PSNR and SSIM with averaged values over 500 reconstruction trails, respectively. We observe that the overall behavior is similar to that of Fig.~\ref{fig: RecPSNR of Simulation1} and~\ref{fig: RecSSIM of Simulation1} except that both the PSNR and SSIM curves of Duarte's method~\cite{duarte2009learning} fluctuate in the lowest level for the whole SR region. Accordingly, Duarte's method~\cite{duarte2009learning} also fails to retrieve the images in Fig.~\ref{fig: RecImg of Simulation2}. This is mainly because Duarte's method~\cite{duarte2009learning} is demanding in the quantization accuracy. When large quantization errors are introduced, sparse coefficients of the test images may not be correctly calculated by the reconstruction algorithm. The PSNR and SSIM curves of the NGI method lie in a low level similar to that of Duarte's method, and the images are only vaguely reconstructed, as shown in Fig.\ref{fig: RecImg of Simulation2}. 
We also observe that the performance of Xu's method~\cite{xu2015optimization} becomes worse in the quantized case, although the images can still be retrieved.    Overall, our proposed method performs the best for both the high accurate as well as the quantized scenarios.

\subsection{Experimental Results} \label{sec:exp}
\par We also experimentally compare the proposed method with the Gaussian method, Duarte's method~\cite{duarte2009learning} and NGI method~\cite{sun2012normalized}. The schematic diagram of experimental setup is shown as Fig.~\ref{fig: ExperimentalSetup}. The light-field patterns are first displayed on the digital micro-mirror device (DMD) after preloaded via the computer. Next, light from a light-emitting diode (LED) source is modulated by a Kohler illumination system to be evenly incident on DMD. 
The light reflected by DMD is then projected onto the imaging object by a lens system. Finally, the whole light reflected from the object is collected by the lens and measured by the detector. 
In our experiments, the light-field patterns are displayed at a rate of 10Hz to avoid frame dropping of the detector, so that the sampling procedure lasts for one minute or so.
After the samplings, the subsequent image retrieval steps for each method are the same as those in the simulation test. 
The reconstruction was carried out on an industrial computer with $32$GB RAM and Intel(R) Core(TM)-I7 $2600$ CPU @$3.4$GHz, and the consuming time of matrix optimization and reconstruction for different methods is specified in Table~\ref{tab:2}. The reconstruction time for each method is given as a time period, since it varies according to the sampling rate.

\begin{table}[t] \label{tab:2}
\caption{Running time of test methods}
\setlength{\arrayrulewidth}{.75pt}
\centering
\begin{tabular}{c|c|c}
\hline
\textbf{Method} & \textbf{Matrix Optimization (sec.)} & \textbf{Reconstruction (sec.)} \\
\hline \hline 
Proposed & 0.365 & $0.037$ to $0.150$ \\
\hline 
Gaussian & -- & $0.037$ to $0.158$ \\
\hline 
Duarte~\cite{duarte2009learning} & $0.345$ & $0.039$ to $0.157$ \\
\hline 
Xu~\cite{xu2015optimization} & $90.18$ (for $100$ iterations) & $0.028$ to $0.360$ \\
\hline 
NGI~\cite{sun2012normalized} & -- & $0.001$ to $0.007$ \\
\hline
\end{tabular}
\label{tab: TimeConsuming}
\end{table}

\par The comparison of reconstructed images by different methods is shown as Fig.~\ref{fig: RecImg of Experiments}, where the ground truth is obtained by pixel-wise detection and serve as a reference image.  By comparing the reconstructed images with the ground truth, again, the PSNR and SSIM are calculated and plotted in Fig.~\ref{fig: RecPSNR of Experiments} and~\ref{fig: RecSSIM of Experiments}, respectively. Overall, the experimental results demonstrate that the reconstructed quality of the proposed optimization scheme is superior to that of other methods under test, which well matches our simulation results.

\subsection{Discussions} \label{sec:dicu}
We would like to point out some interesting points that arise from the simulation and experimental results. 

\begin{figure}[t]
\begin{center}
\subfigure[]{\includegraphics[width=0.85\textwidth]{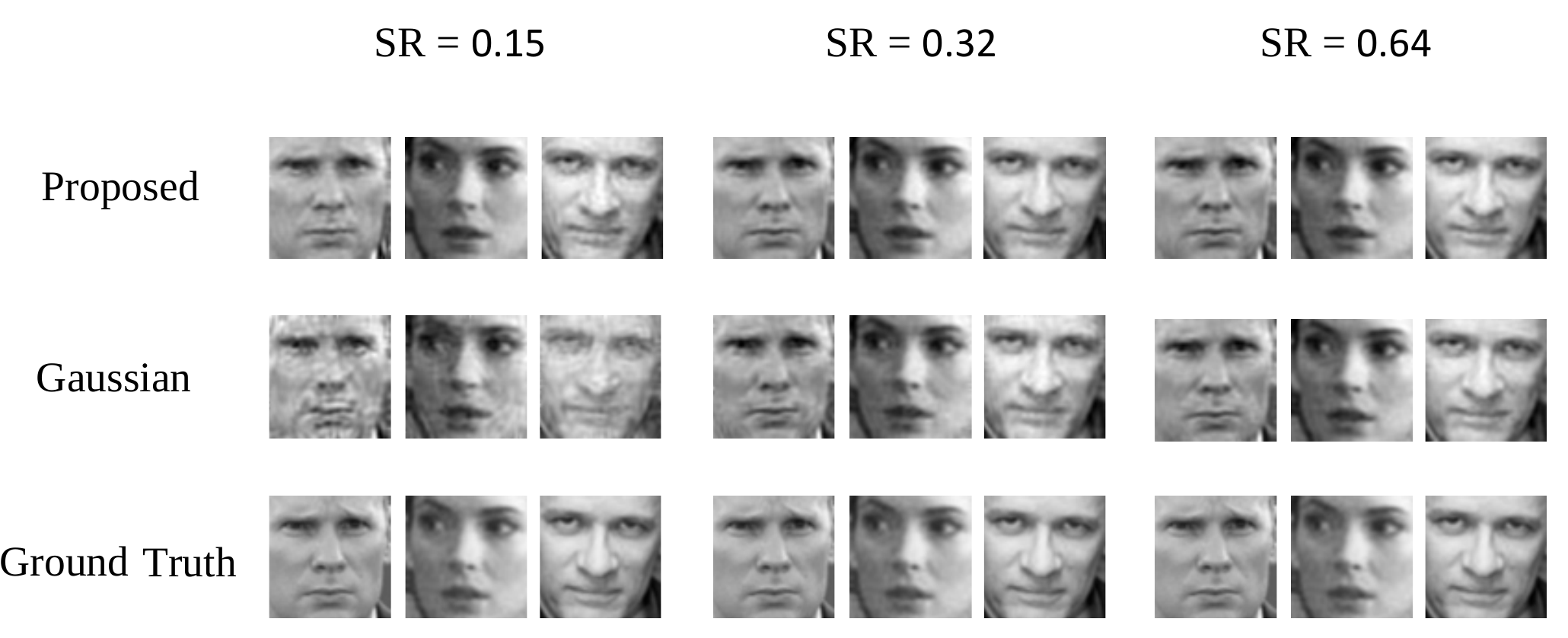}\label{fig:601}} 
\subfigure[]{\includegraphics[width=0.45\textwidth]{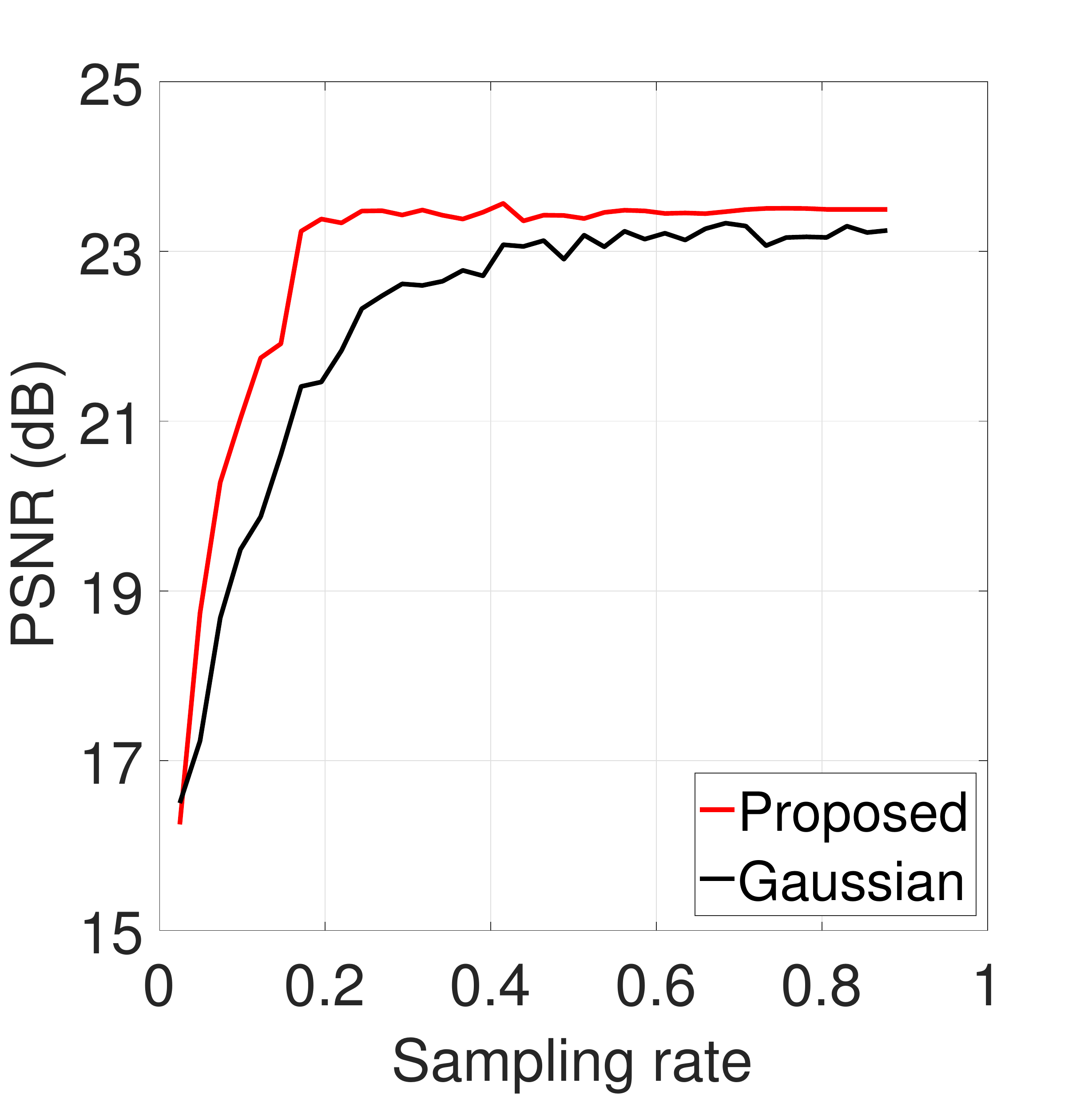} \label{fig:602}} 
\subfigure[]{\includegraphics[width=0.45\textwidth]{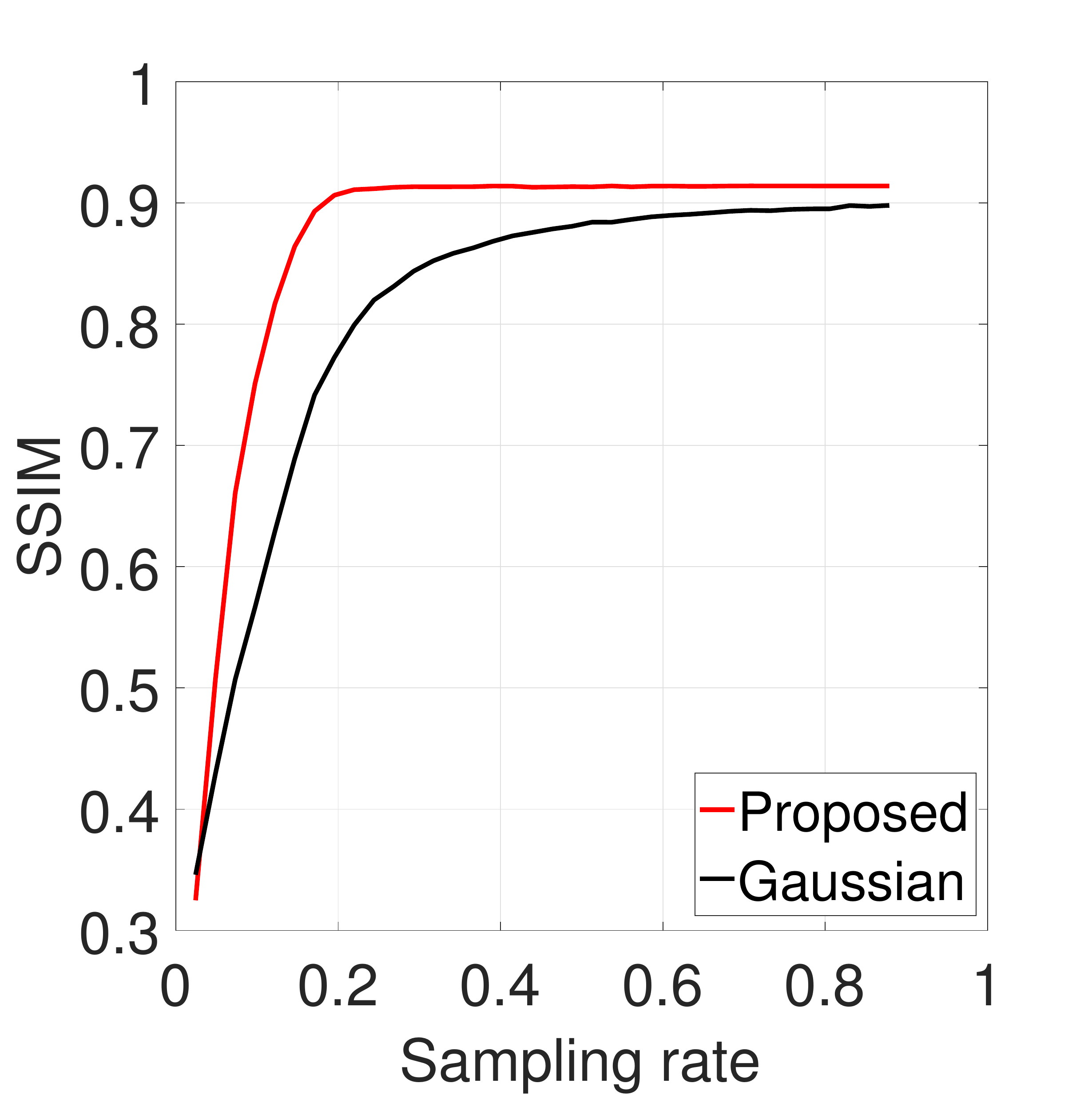}  \label{fig:603}}
\end{center}
\caption{Simulation results of LFWcrop face images. Fig.~\ref{fig:601} shows the reconstructed images via different methods under different SR. Fig.~\ref{fig:602} and \ref{fig:603} show PSNR and SSIM of the reconstructed images via different methods as a function of SR, respectively.} 
\label{fig: Face}
\end{figure}

\begin{itemize}
\item Firstly, the superiority of the proposed method is mainly owing to two factors: i)~optimization of sampling matrix and ii)~dictionary learning. Indeed, the proposed optimization scheme outperforms the Gaussian method, even though they share the same spasifying basis obtained by dictionary learning. This is because our method essentially performs a ``global'' optimization of light fields that incorporates the image statistics captured in the dictionary learning process, thereby enhancing the sampling efficiency. Besides, the improvement of the proposed method over Xu's method can be attributed to the use of both dictionary learning and our optimized sampling matrix.

\item  Secondly, the PSNR and SSIM curves of the proposed method tend towards flat after the SR reaches a critical value. This in turn implies that the inherent information of the imaging object acquired at this very SR value already suffices to produce a satisfactory reconstruction. The critical SR can thus be utilized to evaluate the capability of information acquisition and also allows the comparison of different approaches.

\item 
Thirdly, we would like to point out a practical limitation of the proposed scheme in handling images of large size due to the use of dictionary learning. Specifically,
while dictionary learning in our scheme can bring in some performance gain, it is usually demanding in the requirements of storage and computational cost. Thus the patch size used in dictionary learning should not be large, which, however, poses a limitation to the image size that we can handle. Nevertheless, efficient dictionary learning methods dealing with images of larger scales have recently been proposed~\cite{le2015chasing,sulam2016trainlets,dantas2017learning}, in which the handled image size can go beyond $64 \times 64$ pixels. To demonstrate the effectiveness of the proposed method for images of larger size, we carry out simulations over the LFWcrop database~\cite{sanderson2009multi}, which consists of more than $13,000$ images of $64 \times 64$ pixels. The dictionary is trained offline using the algorithm in~\cite{sulam2016trainlets} with $12,000$ images, which takes about $20$ hours in our industrial computer. The results of the proposed method and the Gaussian method, which involve dictionary learning,  are shown as Fig.~{\ref{fig: Face}} for comparison.

\item Finally, we mention that if one wish to deal with images of even larger size such that existing dictionary learning methods fail to handle or cannot learn the images offline, then the proposed light-field optimization scheme can still be applied by using explicit dictionaries (e.g., Cropped Wavelets~\cite{sulam2016trainlets}) to  incorporate the sparse prior of images.

\end{itemize} 

\section{Conclusion}
In this paper, an optimization scheme of light fields has been proposed to improve the imaging quality of GI. The key idea is to minimize the mutual coherence of the equivalent sampling matrix in order to enhance the sampling efficiency.
A closed-form solution of the sampling matrix has been derived, which enables successive sampling. Simulation and experimental results have shown that the proposed scheme is very effective in improving the reconstruction quality of images, compared to the state-of-the-art methods for GI. The proposed scheme can thus be used to imaging specific targets with higher quality. 
We would also like to point out a technical limitation in our scheme.
Recall that we have employed a NN lifting to cope with the constraint $\mathbf{\Phi}_{ij} \geq 0$. This operation, however, may severely influence the incoherence of the equivalent sampling matrix in the worst case, though such situation rarely happens as confirmed by our empirical test.
Deriving analytical results can better address the non-negative issue while also would require a bit more effort, and our future work will be directed towards this avenue.

\section*{Funding}
National Key Research and Development Program of China (2017YFB0503303, 2017YFB0503300); National Natural Science Foundation of China (NSFC) (11627811).

\bibliography{Ref}

\begin{thebibliography}{10}
\newcommand{\enquote}[1]{``#1''}

\bibitem{pittman1995optical}
T.~B. Pittman, Y.~H. Shih, D.~V. Strekalov, and A.~V. Sergienko,
  \enquote{Optical imaging by means of 2-photon quantum entanglement,}
  {\protect\JournalTitle{Physical Review A Atomic Molecular \& Optical
  Physics}} \textbf{52}, R3429--R3432 (1995).

\bibitem{strekalov1995observation}
D.~Strekalov, A.~Sergienko, D.~Klyshko, and Y.~Shih, \enquote{Observation of
  two-photon ``ghost'' interference and diffraction,}
  {\protect\JournalTitle{Physical Review Letters}} \textbf{74}, 3600 (1995).

\bibitem{bennink2002two}
R.~S. Bennink, S.~J. Bentley, and R.~W. Boyd, \enquote{``two-photon''
  coincidence imaging with a classical source,} {\protect\JournalTitle{Physical
  Review Letters}} \textbf{89}, 113601 (2002).

\bibitem{gatti2004ghost}
A.~Gatti, E.~Brambilla, M.~Bache, and L.~A. Lugiato, \enquote{Ghost imaging
  with thermal light: comparing entanglement and classicalcorrelation,}
  {\protect\JournalTitle{Physical Review Letters}} \textbf{93}, 093602 (2004).

\bibitem{cheng2004incoherent}
J.~Cheng and S.~Han, \enquote{Incoherent coincidence imaging and its
  applicability in x-ray diffraction,} {\protect\JournalTitle{Physical Review
  Letters}} \textbf{92}, 093903 (2004).

\bibitem{zhang2005correlated}
D.~Zhang, Y.-H. Zhai, L.-A. Wu, and X.-H. Chen, \enquote{Correlated two-photon
  imaging with true thermal light,} {\protect\JournalTitle{Optics Letters}}
  \textbf{30}, 2354--2356 (2005).

\bibitem{khakimov2016ghost}
R.~I. Khakimov, B.~Henson, D.~Shin, S.~Hodgman, R.~Dall, K.~Baldwin, and
  A.~Truscott, \enquote{Ghost imaging with atoms,}
  {\protect\JournalTitle{Nature}} \textbf{540}, 100 (2016).

\bibitem{li2018electron}
S.~Li, F.~Cropp, K.~Kabra, T.~Lane, G.~Wetzstein, P.~Musumeci, and D.~Ratner,
  \enquote{Electron ghost imaging,} {\protect\JournalTitle{Physical Review
  Letters}} \textbf{121}, 114801 (2018).

\bibitem{malik2012quantum}
M.~Malik, O.~S. Maga{\~n}a-Loaiza, and R.~W. Boyd, \enquote{Quantum-secured
  imaging,} {\protect\JournalTitle{Applied Physics Letters}} \textbf{101},
  241103 (2012).

\bibitem{zhao2012ghost}
C.~Zhao, W.~Gong, M.~Chen, E.~Li, H.~Wang, W.~Xu, and S.~Han, \enquote{Ghost
  imaging lidar via sparsity constraints,} {\protect\JournalTitle{Applied
  Physics Letters}} \textbf{101}, 141123 (2012).

\bibitem{Gong2011Correlated}
W.~Gong and S.~Han, \enquote{Correlated imaging in scattering media,}
  {\protect\JournalTitle{Optics Letters}} \textbf{36}, 394--6 (2011).

\bibitem{Bina2013Backscattering}
M.~Bina, D.~Magatti, M.~Molteni, A.~Gatti, L.~A. Lugiato, and F.~Ferri,
  \enquote{Backscattering differential ghost imaging in turbid media,}
  {\protect\JournalTitle{Physical Review Letters}} \textbf{110}, 083901 (2013).

\bibitem{wang2016hyperspectral}
Y.~Wang, J.~Suo, J.~Fan, and Q.~Dai, \enquote{Hyperspectral computational ghost
  imaging via temporal multiplexing,} {\protect\JournalTitle{IEEE Photonics
  Technology Letters}} \textbf{28}, 288--291 (2016).

\bibitem{liu2016spectral}
Z.~Liu, S.~Tan, J.~Wu, E.~Li, X.~Shen, and S.~Han, \enquote{Spectral camera
  based on ghost imaging via sparsity constraints,}
  {\protect\JournalTitle{Scientific Reports}} \textbf{6}, 25718 (2016).

\bibitem{morris2015imaging}
P.~A. Morris, R.~S. Aspden, J.~E. Bell, R.~W. Boyd, and M.~J. Padgett,
  \enquote{Imaging with a small number of photons,}
  {\protect\JournalTitle{Nature Communications}} \textbf{6}, 5913 (2015).

\bibitem{liu2018fast}
X.~Liu, J.~Shi, X.~Wu, and G.~Zeng, \enquote{Fast first-photon ghost imaging,}
  {\protect\JournalTitle{Scientific Reports}} \textbf{8}, 5012 (2018).

\bibitem{pelliccia2016experimental}
D.~Pelliccia, A.~Rack, M.~Scheel, V.~Cantelli, and D.~M. Paganin,
  \enquote{Experimental x-ray ghost imaging,} {\protect\JournalTitle{Physical
  Review Letters}} \textbf{117}, 113902 (2016).

\bibitem{yu2016fourier}
H.~Yu, R.~Lu, S.~Han, H.~Xie, G.~Du, T.~Xiao, and D.~Zhu,
  \enquote{Fourier-transform ghost imaging with hard x rays,}
  {\protect\JournalTitle{Physical Review Letters}} \textbf{117}, 113901 (2016).

\bibitem{xia2008experimental}
X.~Shen, Y.~Bai, T.~Qin, and S.~Han, \enquote{Experimental investigation of
  quality of lensless ghost imaging with pseudo-thermal light,}
  {\protect\JournalTitle{Chinese Physics Letters}} \textbf{25}, 3968 (2008).

\bibitem{erkmen2009signal}
B.~I. Erkmen and J.~H. Shapiro, \enquote{Signal-to-noise ratio of
  gaussian-state ghost imaging,} {\protect\JournalTitle{Physical Review A}}
  \textbf{79}, 023833 (2009).

\bibitem{ferri2010differential}
F.~Ferri, D.~Magatti, L.~Lugiato, and A.~Gatti, \enquote{Differential ghost
  imaging,} {\protect\JournalTitle{Physical Review Letters}} \textbf{104},
  253603 (2010).

\bibitem{gong2010method}
W.~Gong and S.~Han, \enquote{A method to improve the visibility of ghost images
  obtained by thermal light,} {\protect\JournalTitle{Physics Letters A}}
  \textbf{374}, 1005--1008 (2010).

\bibitem{brida2010experimental}
G.~Brida, M.~Genovese, and I.~R. Berchera, \enquote{Experimental realization of
  sub-shot-noise quantum imaging,} {\protect\JournalTitle{Nature Photonics}}
  \textbf{4}, 227 (2010).

\bibitem{sun2012normalized}
B.~Sun, S.~S. Welsh, M.~P. Edgar, J.~H. Shapiro, and M.~J. Padgett,
  \enquote{Normalized ghost imaging,} {\protect\JournalTitle{Optics Express}}
  \textbf{20}, 16892--16901 (2012).

\bibitem{Antonini1992Image}
M.~Antonini, M.~Barlaud, P.~Mathieu, and I.~Daubechies, \enquote{Image coding
  using wavelet transform.} {\protect\JournalTitle{IEEE Transactions Image
  Processing}} \textbf{1}, 205--220 (1992).

\bibitem{donoho2006compressed}
D.~L. Donoho, \enquote{Compressed sensing,} {\protect\JournalTitle{IEEE
  Transactions Information Theory}} \textbf{52}, 1289--1306 (2006).

\bibitem{candes2005decoding}
E.~J. Cand{\`e}s and T.~Tao, \enquote{Decoding by linear programming,}
  {\protect\JournalTitle{IEEE Transactions Information Theory}} \textbf{51},
  4203--4215 (2005).

\bibitem{candes2006near}
E.~J. Cand{\`e}s and T.~Tao, \enquote{Near-optimal signal recovery from random
  projections: Universal encoding strategies?} {\protect\JournalTitle{IEEE
  Transactions Information Theory}} \textbf{52}, 5406--5425 (2006).

\bibitem{katz2009compressive}
O.~Katz, Y.~Bromberg, and Y.~Silberberg, \enquote{Compressive ghost imaging,}
  {\protect\JournalTitle{Applied Physics Letters}} \textbf{95}, 131110 (2009).

\bibitem{han2018review}
S.~Han, H.~Yu, X.~Shen, H.~Liu, W.~Gong, and Z.~Liu, \enquote{A review of ghost
  imaging via sparsity constraints,} {\protect\JournalTitle{Applied Sciences}}
  \textbf{8}, 1379 (2018).

\bibitem{gong2013experimental}
W.~Gong, Z.~Bo, E.~Li, and S.~Han, \enquote{Experimental investigation of the
  quality of ghost imaging via sparsity constraints,}
  {\protect\JournalTitle{Applied Optics}} \textbf{52}, 3510--3515 (2013).

\bibitem{chen2014application}
M.~Chen, E.~Li, and S.~Han, \enquote{Application of multi-correlation-scale
  measurement matrices in ghost imaging via sparsity constraints,}
  {\protect\JournalTitle{Applied Optics}} \textbf{53}, 2924--2928 (2014).

\bibitem{khamoushi2015sinusoidal}
S.~M. Khamoushi, Y.~Nosrati, and S.~H. Tavassoli, \enquote{Sinusoidal ghost
  imaging,} {\protect\JournalTitle{Optics Letters}} \textbf{40}, 3452--3455
  (2015).

\bibitem{Li2013Mutual}
E.~Li, M.~Chen, W.~Gong, H.~Yu, and S.~Han, \enquote{Mutual information of
  ghost imaging systems,} {\protect\JournalTitle{Acta Optica Sinica}}
  \textbf{33}, 93--98 (2013).

\bibitem{xu2015optimization}
X.~Xu, E.~Li, X.~Shen, and S.~Han, \enquote{Optimization of speckle patterns in
  ghost imaging via sparse constraints by mutual coherence minimization,}
  {\protect\JournalTitle{Chinese Optics Letters}} \textbf{13}, 071101 (2015).

\bibitem{olshausen1996natural}
B.~A. Olshausen and D.~J. Field, \enquote{Natural image statistics and
  efficient coding,} {\protect\JournalTitle{Network: Computation in Neural
  Systems}} \textbf{7}, 333--339 (1996).

\bibitem{aharon2006rm}
M.~Aharon, M.~Elad, and A.~Bruckstein, \enquote{K-svd: An algorithm for
  designing overcomplete dictionaries for sparse representation,}
  {\protect\JournalTitle{IEEE Transactions Signal Processing}} \textbf{54},
  4311--4322 (2006).

\bibitem{elad2007optimized}
M.~Elad, \enquote{Optimized projections for compressed sensing,}
  {\protect\JournalTitle{IEEE Transactions Signal Processing}} \textbf{55},
  5695--5702 (2007).

\bibitem{abolghasemi2012gradient}
V.~Abolghasemi, S.~Ferdowsi, and S.~Sanei, \enquote{A gradient-based
  alternating minimization approach for optimization of the measurement matrix
  in compressive sensing,} {\protect\JournalTitle{Signal Processing}}
  \textbf{92}, 999--1009 (2012).

\bibitem{duarte2009learning}
J.~M. Duarte-Carvajalino and G.~Sapiro, \enquote{Learning to sense sparse
  signals: Simultaneous sensing matrix and sparsifying dictionary
  optimization,} {\protect\JournalTitle{IEEE Transactions Image Processing}}
  \textbf{18}, 1395--1408 (2009).

\bibitem{donoho2003optimally}
D.~L. Donoho and M.~Elad, \enquote{Optimally sparse representation in general
  (nonorthogonal) dictionaries via $l 1$ minimization,}
  {\protect\JournalTitle{Proceedings of the National Academy of Sciences}}
  \textbf{100}, 2197--2202 (2003).

\bibitem{pati1993orthogonal}
Y.~C. {Pati}, R.~{Rezaiifar}, and P.~S. {Krishnaprasad}, \enquote{Orthogonal
  matching pursuit: recursive function approximation with applications to
  wavelet decomposition,} in \emph{Proceedings of 27th Asilomar Conference on
  Signals, Systems and Computers,}  (IEEE, 1993), pp. 40--44.

\bibitem{tropp2004greed}
J.~A. Tropp, \enquote{Greed is good: Algorithmic results for sparse
  approximation,} {\protect\JournalTitle{IEEE Transactions Information Theory}}
  \textbf{50}, 2231--2242 (2004).

\bibitem{cleju2014optimized}
N.~Cleju, \enquote{Optimized projections for compressed sensing via
  rank-constrained nearest correlation matrix,} {\protect\JournalTitle{Applied
  and Computational Harmonic Analysis}} \textbf{36}, 495--507 (2014).

\bibitem{deng2012mnist}
L.~Deng, \enquote{The mnist database of handwritten digit images for machine
  learning research [best of the web],} {\protect\JournalTitle{IEEE Signal
  Processing Magazine}} \textbf{29}, 141--142 (2012).

\bibitem{le2015chasing}
L.~Le~Magoarou and R.~Gribonval, \enquote{Chasing butterflies: In search of
  efficient dictionaries,} in \emph{2015 IEEE International Conference on
  Acoustics, Speech and Signal Processing (ICASSP),}  (IEEE, 2015), pp.
  3287--3291.

\bibitem{sulam2016trainlets}
J.~Sulam, B.~Ophir, M.~Zibulevsky, and M.~Elad, \enquote{Trainlets: Dictionary
  learning in high dimensions,} {\protect\JournalTitle{IEEE Transactions Signal
  Processing}} \textbf{64}, 3180--3193 (2016).

\bibitem{dantas2017learning}
C.~F. Dantas, M.~N. Da~Costa, and R.~da~Rocha~Lopes, \enquote{Learning
  dictionaries as a sum of kronecker products,} {\protect\JournalTitle{IEEE
  Signal Processing Letters}} \textbf{24}, 559--563 (2017).

\bibitem{sanderson2009multi}
C.~Sanderson and B.~C. Lovell, \enquote{Multi-region probabilistic histograms
  for robust and scalable identity inference,} in \emph{International
  Conference on Biometrics,}  (Springer, 2009), pp. 199--208.

\end{thebibliography}

\clearpage

\end{document}